\documentclass[prd,aps,showpacs,nofootinbib,eqsecnum,onecolumn,groupedaddress,
amssymb]{revtex4}
\usepackage{amsmath} 
\usepackage{amssymb}
\usepackage{amsfonts}
\usepackage{graphicx,bm}
\usepackage{dcolumn}
\usepackage{color,amsxtra}
\usepackage{epsf}
\usepackage{enumerate}
\usepackage{hhline}
\usepackage{array}
\usepackage{tabularx}
\usepackage{subfigure}

\usepackage{fancyhdr}
\usepackage{mathrsfs}

\newcommand{\be}{\begin{equation}}
\newcommand{\ee}{\end{equation}}
\newcommand{\bea}{\begin{eqnarray}}
\newcommand{\eea}{\end{eqnarray}}
\newcommand{\beaa}{\begin{eqnarray*}}
\newcommand{\eeaa}{\end{eqnarray*}}

\newcommand{\nn}{\nonumber \\}
\newcommand{\e}{\mathrm{e}}

\allowdisplaybreaks[4]

\newcommand{\Eqn}[1]{&\hspace{-0.2em}#1\hspace{-0.2em}&}

\def\be{\begin{equation}}
\def\ee{\end{equation}}
\def\bea{\begin{eqnarray}}
\def\eea{\end{eqnarray}}

\def\nn{\nonumber \\}
\def\e{\mathrm{e}}

\begin{document}

\title{Inflationary universe from perfect fluid and $F(R)$ gravity and its 
comparison with observational data} 

\author{Kazuharu Bamba$^{1, 2}$, Shin'ichi Nojiri$^{3, 4}$, 
Sergei D. Odintsov$^{5, 6}$ and Diego S\'{a}ez-G\'{o}mez$^{7, 8}$
}
\affiliation{
$^1$Leading Graduate School Promotion Center,
Ochanomizu University, 2-1-1 Ohtsuka, Bunkyo-ku, Tokyo 112-8610, Japan\\
$^2$Department of Physics, Graduate School of Humanities and Sciences, Ochanomizu University, Tokyo 112-8610, Japan\\
$^3$Department of Physics, Nagoya University, Nagoya 464-8602, Japan\\
$^4$Kobayashi-Maskawa Institute for the Origin of Particles and the
Universe, Nagoya University, Nagoya 464-8602, Japan\\
$^5$Consejo Superior de Investigaciones Cient\'{\i}ficas, ICE/CSIC-IEEC, 
Campus UAB, Facultat de Ci\`{e}ncies, Torre C5-Parell-2a pl, E-08193
Bellaterra (Barcelona), Spain\\
$^6$Instituci\'{o} Catalana de Recerca i Estudis Avan\c{c}ats
(ICREA), Barcelona, Spain\\ 
$^7$Astrophysics, Cosmology and Gravity Centre (ACGC) and \\ 
Department of Mathematics and Applied Mathematics, University of Cape Town, Rondebosch 7701, Cape Town, South Africa \\
$^8$Fisika Teorikoaren eta Zientziaren Historia Saila, Zientzia eta Teknologia Fakultatea,\\
Euskal Herriko Unibertsitatea, 644 Posta Kutxatila, 48080 Bilbao, Spain
}


\begin{abstract} 
We investigate the descriptions for the observables of inflationary 
models, in particular, the spectral index of curvature perturbations, 
the tensor-to-scalar ratio, and the running of the spectral index, 
in the framework of perfect fluid models and $F(R)$ gravity theories 
through the reconstruction methods. 
Furthermore, the perfect fluid and $F(R)$ gravity descriptions 
of inflation are compared with the recent cosmological observations 
such as the Planck satellite and BICEP2 experiment. 
It is demonstrated with explicit examples that perfect fluid may lead to the 
inflationary universe consistent with the Planck data. It is also shown 
that several $F(R)$ gravity models, especially, a power-law model 
gives the best fit values compatible with the spectral index and tensor-to-scalar ratio within the allowed ranges suggested by the Planck and BICEP2 results. 
\end{abstract}

\pacs{04.50.Kd, 98.80.-k, 98.80.Cq, 12.60.-i}
\hspace{14.5cm} OCHA-PP-328

\maketitle

\def\thesection{\Roman{section}}
\def\theequation{\Roman{section}.\arabic{equation}}

\section{Introduction}

Owing to the recent data taken by the BICEP2 experiment~\cite{Ade:2014xna} 
on the tensor-to-scalar ratio of the primordial density perturbations, 
additionally to the observations by the satellites of 
the Wilkinson Microwave anisotropy probe (WMAP)~\cite{WMAP, Hinshaw:2012aka} and the Planck~\cite{Ade:2013lta, Ade:2013uln}, 
inflation has attracted much more attention. 

The potential form of inflaton is related to 
the spectrum of the density perturbations generated during inflation~\cite{Lidsey:1995np}. Several models of inflation have recently been constructed to 
account for the Planck and BICEP2 data~\cite{HSSS-WLLQCZ}. 
Particularly, scalar field models of inflation have been explored 
in comparison with the data analysis of the BICEP2~\cite{Scalar-BICEP2, Gong:2014cqa}. 

Recently, in Ref.~\cite{Bamba:2014daa}, we have explicitly performed the reconstruction of scalar field theories with inflation leading to the theoretical consequences compatible with the observational data obtained from the Planck and BICEP2 in terms of the spectral index of the curvature fluctuations, the tensor-to-scalar ratio, and the running of the spectral index. 
As the developments of these investigations, in this paper, we reconstruct 
the descriptions of inflation in 
perfect fluid models and $F(R)$ gravity theories\footnote{For recent reviews on dark energy problem and modified gravity theories including $F(R)$ gravity to solve it, see, for instance, Refs.~\cite{R-NO-CF-CD, Bamba:2012cp, Joyce:2014kja}.}. 
Especially, we re-express the observables of inflationary models, i.e., 
the spectral index of curvature perturbations, 
the tensor-to-scalar ratio, and the running of the spectral index, 
in terms of the quantities in perfect fluid models and 
$F(R)$ gravity theories. 
We also compare the perfect fluid and $F(R)$ gravity descriptions 
of inflation with the recent observational data obtained from 
the Planck satellite and BICEP2 experiment. 
An inflationary model of the fluid with its inhomogeneous viscosity 
has recently been discussed in Ref.~\cite{Myrzakulov:2014kta}. 
Such reconstructions also work to realize bounce 
universes~\cite{Prof-Brandenberger, Bamba:2013fha}. 
The reconstruction of $F(R)$ gravity model from observational data has 
been executed in Ref.~\cite{Rinaldi:2014gua}.  
We note that the first successful inflationary model in such a sort 
of modified gravity (i.e., $R^2$ inflation) has been proposed 
in Ref.~\cite{staro}. There have also been existed its supergravity extension~\cite{FKR-FP} and the related models~\cite{SCMOZ-CM}. 
We remark that there have been the attempts to make modified gravity models to explain the Planck and BICEP2 
results~\cite{Bamba:2014jia, BCOZ-CEZ, LT-TW}. 
Moreover, the features of the spectral index 
in induced gravity~\cite{Kaiser:1994wj} and scalar-tensor theories~\cite{Prof-Kaiser} have been examined in detail. 

We here emphasize the physical motivations of this work and explain 
the importance of formulations of the observables for 
inflationary models in terms of the perfect fluid and $F(R)$ gravity. 
In ordinary scalar field model of inflation, the 
the spectral index, the tensor-to-scalar ratio, and the running of the spectral index are represented by using the potential of the scalar field. 
Therefore, we can judge what scalar field models 
are consistent with the observations. 
Similarly, if we re-formulate these observables for 
inflationary models in the perfect fluid and $F(R)$ gravity 
descriptions, by comparing these theoretical representations 
with the observations, we can obtain information 
on the properties of the perfect fluid and $F(R)$ gravity models to 
account for the observations in the early universe. 
Thus, the outcome of this approach is that 
since the natures of the perfect fluid and $F(R)$ gravity models 
have been examined by using the observations about 
the current cosmic acceleration, by combining the 
insights obtained from our approach in terms of inflation, 
we can find novel conditions for the perfect fluid and $F(R)$ gravity models 
to be viable from the cosmological point of view. 
We use units of $k_\mathrm{B} = c = \hbar = 1$ and express the
gravitational constant $8 \pi G_\mathrm{N}$ by
${\kappa}^2 \equiv 8\pi/{M_{\mathrm{Pl}}}^2$ 
with the Planck mass of $M_{\mathrm{Pl}} = G_\mathrm{N}^{-1/2} = 1.2 \times 
10^{19}$\,\,GeV. 

The organization of the paper is the following. 
In Sec.~II, we write the slow-roll parameters and express the 
observables of inflationary models in the formulation of the perfect fluid. 
Next, in Sec.~III we represent the slow-roll parameters and 
the observables of inflationary models with the quantities in $F(R)$ gravity. 
These descriptions are also compared with the Planck and BICEP2 data. 
Conclusions are presented in Sec.~IV. 
The explicit expressions of the observables for inflationary models 
in the description of the perfect fluid 
and those in the $F(R)$ gravity description 
are given in Appendixes A and B, respectively. 
In Appendix C, we present the representations of the observables for 
inflationary models in the linear form of square of the Hubble parameter 
and those in the exponential one. 
In Appendix D, the relation between the equation of state (EoS) parameter and the tensor-to-scalar ratio is described. 

\section{Perfect fluid description of the slow-roll parameters} 

Usually, the slow-roll parameters are related to the potential of the scalar field, namely, inflaton. In this section, we explore how these parameters may be expressed in terms of perfect fluid. 

\subsection{Slow-roll parameters} 

For the model of a scalar field $\phi$ coupled with gravity 
\be
\label{S1}
S = \int d^4 x \sqrt{-g} \left( \frac{R}{2\kappa^2} 
 - \frac{1}{2}\partial_\mu \phi \partial^\mu \phi - V(\phi) \right)\, ,
\ee
where $R$ is the scalar curvature, 
we define the slow-roll parameters $\epsilon$, $\eta$ and $\xi$ as follows, 
\be
\label{S2}
\epsilon \equiv 
\frac{1}{2\kappa^2} \left( \frac{V'(\phi)}{V(\phi)} \right)^2\, , 
\quad 
\eta \equiv \frac{1}{\kappa^2} \frac{V''(\phi)}{V(\phi)}\, , 
\quad 
\xi^2 \equiv \frac{1}{\kappa^4} \frac{V'(\phi) V'''(\phi)}{\left(V(\phi)\right)^2}\, . 
\ee 
Here and in the following, 
the prime means the derivative with respect to the argument such as 
$V' (\phi) \equiv \partial V(\phi)/\partial \phi$, etc. 
For the scalar model, we find that 
the spectral index $n_\mathrm{s}$ of the curvature perturbations,  
the tensor-to-scalar ratio $r$ of the density perturbations, 
and the running of the spectral index $\alpha_\mathrm{s}$ can be expressed as 
\be
\label{S3}
n_\mathrm{s} - 1 \sim - 6 \epsilon + 2 \eta\, , 
\quad 
r = 16 \epsilon \, , 
\quad 
\alpha_\mathrm{s} \equiv \frac{d n_\mathrm{s}}{d \ln k} 
\sim 16\epsilon \eta - 24 \epsilon^2 - 2 \xi^2\, .  
\ee
We suppose the flat Friedmann-Lema\^{i}tre-Robertson-Walker (FLRW) metric 
\be
\label{FRW}
ds^2 = -dt^2 + a^2(t) \sum_{i=1,2,3}\left(dx^i\right)^2 \, .
\ee
Here, $a(t)$ is the scale factor and the Hubble ratio is  
defined as $H \equiv \dot{a}/a$ with the dot $\dot\ $ expressing 
the derivative with respect to time, $\partial/\partial t$. 

We describe the expressions of the slow-roll parameters with $H$. 
For the action in Eq.~(\ref{S1}), the gravitational equations in the FLRW background in Eq.~(\ref{FRW}) are given by 
\bea
\label{S4}
\frac{3}{\kappa^2} H^2 \Eqn{=} \frac{1}{2}{\dot \phi}^2 + V(\phi)\, ,  \\
\label{S4-2} 
- \frac{1}{\kappa^2} \left( 3 H^2 + 2\dot H \right) 
\Eqn{=} \frac{1}{2}{\dot \phi}^2 - V(\phi)\, . 
\eea
We now redefine the scalar field $\phi$ by a new scalar field $\varphi$, 
$\phi = \phi (\varphi)$ and identify $\varphi$ with the number of 
$e$-folds $N$. 
Then, Eqs.~(\ref{S4}) and (\ref{S4-2}) can be rewritten as follows, 
\bea
\label{S5}
\frac{3}{\kappa^2} \left(H(N)\right)^2 \Eqn{=} \frac{1}{2}\omega(\varphi) \left(H(N)\right)^2 
+ V\left(\phi\left( \varphi \right) \right)\, , \\
\label{S5-2}
 - \frac{1}{\kappa^2} \left( 3 \left(H(N)\right)^2 + 2H'(N) H(N) \right) 
\Eqn{=} \frac{1}{2}\omega(\varphi) \left(H(N)\right)^2 - V\left(\phi\left( \varphi \right) \right)\, .
\eea
Here, $\omega(\varphi) \equiv \left( d\phi/d\varphi \right)^2$. 
Therefore, if the Hubble expansion rate $H$ is given by a function of $N$ as $H(N)$ and $\omega(\varphi)$ and 
$V(\varphi)\equiv V\left(\phi\left(\varphi\right)\right)$ are written 
in the following form, 
\be
\label{S6}
\omega(\varphi) 
= - \left. \frac{2 H'(N)}{\kappa^2 H(N)}\right|_{N=\varphi}\, , 
\quad 
V(\varphi) = \left. 
\frac{1}{\kappa^2}\left[ 3 \left(H(N)\right)^2 + H(N) H'(N) \right]
\right|_{N=\varphi}\, ,
\ee
we find $H=H(N)$, $\varphi = N$ as a solution for the field equation 
$\phi$ or $\varphi$ and the Einstein equation. 
It should be noted that $H'<0$ because $\omega(\varphi)>0$. 
Thus, we can express the slow-roll parameters $\epsilon$, $\eta$ and $\xi$ 
in terms of $H$. 
The representations are described in (2.8) of Ref.~\cite{Bamba:2014daa}.

\subsection{Perfect fluid description} 

We now rewrite the expressions of the slow-roll 
parameters in terms of the perfect fluid. 
In the FLRW space-time (\ref{FRW}), 
the gravitational field equations for the perfect fluid 
has the following form
\be
\label{SSS1}
\frac{3}{\kappa^2} \left(H (N)\right)^2 = \rho\, ,\quad 
 - \frac{2}{\kappa^2} H(N) H'(N) = \rho + P\, , 
\ee
where $\rho$ and $P$ are the energy density and pressure of the perfect fluid, 
respectively. 
We assume a rather general equation of state (EoS) 
\be
\label{SS2}
P(N) = - \rho(N) + f(\rho)\, ,
\ee
with $f(\rho)$ an arbitrary function of $\rho$. 
In this case, the second equation in (\ref{SSS1}) has the following form:
\be
\label{SS3}
 - \frac{2}{\kappa^2} H(N) H'(N) = f(\rho)\, ,
\ee
and the conservation law $0=\rho'(N) + 3 \left(\rho(N) + P(N)\right)$ reads 
\be
\label{SS4}
0 = \rho'(N) + 3 f(\rho)\, .
\ee
By using Eqs.~(\ref{SS3}) and (\ref{SS4}), we obtain 
\be
\label{SS5}
\frac{2}{\kappa^2} \left[ \left(H'(N)\right)^2 + H(N) H''(N) \right] 
= 3 f'(\rho) f(\rho)\, .
\ee
Here, it is emphasized that the prime operating $f(\rho)$ means the derivative 
with respect to $\rho$ of $f'(\rho) \equiv df(\rho)/d\rho$, 
whereas that $H'(N) \equiv dH(N)/dN$ and $\rho'(N) \equiv d\rho(N)/dN$. 
All the slow-roll parameters can 
be rewritten only in terms of $\rho(N)$ and $f(\rho(N))$. 
Hence, we obtain the expressions of observables for inflationary 
models, $n_\mathrm{s}$, $r$, and $\alpha_\mathrm{s}$, shown in Appendix A.

\subsection{Reconstruction of perfect fluid models}

As examples of the form for the square of the Hubble parameter, 
we examine two models. One is the linear form and the other is 
the exponential one. 

\subsubsection{Linear form} 

As the first example, 
we examine the following linear form for $H^2$:  
\be
\left(H (N)\right)^2=G_0 N + G_1\, ,
\label{eq:IIC1} 
\ee
with $G_0 (<0)$ and $G_1 (>0)$ constants.  

The motivations why we take such a linear form is as follows. 
For the slow-roll exponential inflation, 
the scale factor is given by $a = \bar{a} \exp\left(H_\mathrm{inf} t\right)$ 
with $\bar{a}$ a constant. 
Here, $H_\mathrm{inf}$ is the Hubble parameter at the inflationary stage 
is approximately equal to a constant, namely, its time dependence 
is very weak. To express the weak time dependence of $H$, 
we use the form in Eq.~(\ref{eq:IIC1}), in which the number of $e$-folds 
is considered to play a role of time. 
In this case, if $G_1/G_0 \ll N$, the time dependence 
of the Hubble parameter during inflation is negligible. 

Using the gravitational equations in (\ref{SSS1}) with Eq.~(\ref{eq:IIC1}) 
and $H>0$, we obtain
\be
\rho(N) = \frac{3}{\kappa^2} \left(G_0 N + G_1 \right)\,, 
\quad 
P(N) = -\frac{1}{\kappa^2} \left[ \left(3N + 1 \right)G_0 + 3G_1 \right]
\,.
\label{eq:IIC1-01}
\ee
By eliminating $N$ from these equations, we acquire 
\be
P(N) = -\rho(N) -\frac{G_0}{\kappa^2} \,.
\label{eq:IIC1-03} 
\ee
It follows from Eqs.~(\ref{SS2}) and (\ref{eq:IIC1-03}) that 
\be
f(\rho) = -\frac{G_0}{\kappa^2}\, .
\ee
Therefore, with Eqs.~(\ref{SS2}) and the equations in (\ref{eq:IIC1-01}), 
we see that the EoS reads
\be
w(N) \equiv \frac{P(N)}{\rho(N)} = -1+ \frac{f(\rho)}{\rho(N)} 
= -\frac{\left(3N + 1 \right)G_0 + 3G_1}{3\left(G_0 N + G_1 \right)}\,.
\label{eq:IIC4}  
\ee
%

\subsubsection{Exponential form} 

As the second example, we study the following exponential form for $H^2$: 
\be
\left(H (N)\right)^2=G_2 \e^{\beta N} + G_3\, ,
\label{eq:IIC5} 
\ee
with $G_2 (<0)$, $G_3 (>0)$, and $\beta (>0)$ constants. 

The physical reason why we examine this exponential form   
is the following. 
In the power-law inflation, 
the scale factor is expressed as $a = \bar{a} t^{\hat{p}}$ 
with $\hat{p}$ a constant. 
In this case, the square of the 
the Hubble parameter during inflation becomes 
$H^2 = \left( \hat{p}/t \right)^2 = \hat{p}^2\exp\left(-2N/\hat{p}\right)$. 
This is equivalent to the form in Eq.~(\ref{eq:IIC5}) 
with $G_2 = \hat{p}^2$, $\beta= -2/\hat{p}$, and $G_3 = 0$. 
Thus, such an exponential form can describe the power-law inflation. 

{}From the gravitational equations in (\ref{SSS1}) with Eq.~(\ref{eq:IIC5}) 
and $H>0$, we have
\be
\rho(N) = \frac{3}{\kappa^2} \left(G_2 \e^{\beta N} + G_3 \right)\,, 
\quad 
P(N) = -\frac{1}{\kappa^2} \left[ \left(3 + \beta \right)G_2 \e^{\beta N} + 3G_3 \right]\,. 
\label{eq:IIC1-04}  
\ee
The elimination of $N$ from these equations gives 
\be
P(N) = -\left(1 + \frac{\beta}{3}\right) \rho(N) 
+ \frac{G_3 \beta}{\kappa^2} \,.
\label{eq:IIC1-06} 
\ee
By using the first equation in~(\ref{eq:IIC1-04}) and comparing Eq.~(\ref{eq:IIC1-06}) with Eq.~(\ref{SS2}), we get
\be
f(\rho) = -\frac{\beta}{3} \rho(N) + \frac{G_3 \beta}{\kappa^2} \,. 
\label{eq:IIC1-07} 
\ee
Accordingly, from the first equality in (\ref{eq:IIC4}) we find  
\be
w(N) = - \frac{\left(3 + \beta \right)G_2 \e^{\beta N} + 3G_3}{3\left( G_2 \e^{\beta N} + G_3 \right)} \,. 
\label{eq:IIC1-08}  
\ee
%

\subsubsection{Another form} 

We here mention that 
various models of $H(N)$ or the representations by the equation of 
state have been investigated in Refs.~\cite{Mukhanov:2013tua, Garcia-Bellido:2014gna, Mukhanov:2014uwa}. 
As an example, 
we show 
the following EoS parameter \cite{Mukhanov:2013tua}, 
\be
w=-1+\frac{\tilde{\beta}}{(1+N)^{\tilde{\gamma}}}\,.
\label{N1.1}
\ee
Here, $\{\tilde{\beta},\tilde{\gamma}\}$ are free parameters. Then, by solving the gravitational field equations in the FLRW background, 
the Hubble parameter becomes 
\be
H=\bar{H}\exp\left[-\frac{3\tilde{\beta} (1+N)^{1-\tilde{\gamma}}}{2(1-\tilde{\gamma})} \right] \,, 
\label{N1.2}
\ee
with $\bar{H}$ a constant, 
while the spectral index and the tensor-to-scalar ratio are given by 
\bea
\hspace{-7mm}
n_\mathrm{s}-1 \Eqn{=} \frac{1}{6} \left\{-18 \tilde{\beta}  (N+1)^{-\tilde{\gamma}}-\frac{4 \tilde{\gamma}^2 (N+1)^{\tilde{\gamma} -2} \left[\tilde{\beta} +(N+1)^{\tilde{\gamma} }\right]}{\left[\tilde{\beta} -2 (N+1)^{\tilde{\gamma} }\right]^2}+\frac{6 \tilde{\gamma}  \tilde{\beta}  (N+1)+4 \tilde{\gamma}  (3 N+2) (N+1)^{\tilde{\gamma} }}{(N+1)^2 \left[2 (N+1)^{\tilde{\gamma} }-\tilde{\beta} \right]}\right\}\,,
\label{N1.3} \\
\hspace{-7mm}
r \Eqn{=} \frac{8 \tilde{\beta} (N+1)^{-\tilde{\gamma} -2} \left[\left(
\tilde{\gamma} -6 N-6\right) (N+1)^{\tilde{\gamma} }+3 \tilde{\beta}  (N+1)\right]^2}{3 \left[\tilde{\beta} -2 (N+1)^{\tilde{\gamma} }\right]^2}\,.
\label{N1.3}
\eea
Hence, for the appropriate values of the free parameters $\{\tilde{\beta},\tilde{\gamma}\}$, 
the observational values such as $n_\mathrm{s}$ and $r$ 
can be reproduced. 
We note that the explicit expression of $n_\mathrm{s}$ is also 
presented in Ref.~\cite{Mukhanov:2013tua}.

\subsubsection{Comparison with the observations} 

Provided that the time variation of $f(\rho)$ and $\rho$ during inflation 
is sufficiently small, and that inflation is almost exponential as
$w(N) \equiv P(N)/\rho(N) = -1+ f(\rho)/\rho(N) \approx -1$, i.e., 
$\left| f(\rho)/\rho(N) \right| \ll 1$, 
from the expressions in Appendix A 
we acquire 
\begin{equation} 
n_\mathrm{s} \sim 1 - 6\frac{f(\rho)}{\rho(N)}\,,  
\quad 
r \approx 24 \frac{f(\rho)}{\rho(N)}\,, 
\quad 
\alpha_\mathrm{s} \approx -9 \left(\frac{f(\rho)}{\rho(N)}\right)^2 \,.
\label{eq:2.37} 
\end{equation}

We here present the recent observations on the spectral index 
$n_{\mathrm{s}}$, the tensor-to-scalar ratio $r$, and the running of the spectral index $\alpha_\mathrm{s}$. 
The Planck data~\cite{Ade:2013lta, Ade:2013uln} suggest 
$n_{\mathrm{s}} = 0.9603 \pm 0.0073\, (68\%\,\mathrm{CL})$, 
$r< 0.11\, (95\%\,\mathrm{CL})$, 
and $\alpha_\mathrm{s} = -0.0134 \pm 0.0090\, (68\%\,\mathrm{CL})$ 
[the Planck and WMAP~\cite{WMAP, Hinshaw:2012aka}], the negative sign of 
which is at $1.5 \sigma$. 
The BICEP2 experiment~\cite{Ade:2014xna} implies 
$r = 0.20_{-0.05}^{+0.07}\, (68\%\,\mathrm{CL})$. 
It is mentioned that 
the discussions exist on how to subtract the foreground, 
for example, in Refs.~\cite{A-A, Mortonson:2014bja, Kamionkowski:2014}. 
Recently, there have also appeared the progresses to ensure the 
BICEP2 statements in Ref.~\cite{Colley:2014nna}. 
It is also remarked that the representation of $\alpha_\mathrm{s}$ is also 
given in Ref.~\cite{Bassett:2005xm}. 

It is seen from the equations in (\ref{eq:2.37}) that 
when the condition $f(\rho)/\rho(N) = 6.65 \times 10^{-3}$ 
is realized in the inflationary era, 
we find $(n_{\mathrm{s}}, r, \alpha_\mathrm{s}) = 
(0.960, 0.160, -3.98 \times 10^{-4})$. 
In the linear form of $H^2$ in Eq.~(\ref{eq:IIC1}), 
if $G_1/G_0 \gg N$ 
and $-G_0/\left( 3G_1 \right) = 6.65 \times 10^{-3}$, 
the change of value of $w(N)$ is considered to be negligible, 
and the above condition can be met at the inflationary stage. 
On the other hand, for the exponential form of $H^2$ in Eq.~(\ref{eq:IIC5}), 
provided that $\beta = 2.0 \times 10^{-4}$ and hence $\beta N \ll 1$, 
and that 
$-1/3\left\{\left[1+G_3/\left(G_2 \beta \right)\right]\right\} 
= 6.65 \times 10^{-3}$, then $w(N)$ can be regarded as 
constant and the above condition can be satisfied during inflation. 
As a consequence, it is interpreted that perfect fluid models can 
lead to the Planck result with $r=\mathcal{O}(0.1)$, the value of which 
is compatible with the BICEP2 experiment. 

Moreover, we mention concrete models~\cite{Astashenok:2012kb}. 
For a model of $P(N) = -\rho(N) +f(\rho)$ 
with $f(\rho) = \bar{f} \sin \left( \rho(N)/\bar{\rho} \right)$, 
where $\bar{f}$ is a constant and $\bar{\rho}$ is a fiducial value of $\rho$, 
known to produce the Pseudo-Rip scenario~\cite{P-R-S}, 
if $\rho(N)/\bar{\rho} \ll 1$ and 
$f(\rho)/\rho(N) \approx \bar{f}/\bar{\rho} = 6.65 \times 10^{-3}$, 
$f(\rho)/\rho(N)$ behaves almost constant and 
the above condition can be met. 
In addition, we examine the other model of $P(N) = -\rho(N) +f(\rho)$, 
where $f(\rho) = \left(\rho(N)\right)^{\tau}$ with $\tau (\neq)$ a constant. 
By using the first equation in (\ref{SSS1}), we obtain
$f(\rho)/\rho(N) \approx \left(3H_\mathrm{inf}^2/\kappa^2 \right)^{\tau-1}$, 
where $H_\mathrm{inf} = \mathrm{constant}$ is the Hubble parameter 
at the slow-roll inflation regime. 
Thus, similarly to the first example, when 
$f(\rho)/\rho(N) \approx \left(3H_\mathrm{inf}^2/\kappa^2 \right)^{\tau-1} = 6.65 \times 10^{-3}$, 
$f(\rho)/\rho(N)$ can be considered to be constant and 
the above condition can be satisfied.

\section{$F(R)$ gravity description of the slow-roll parameters} 

Next, we describe the slow-roll parameters in terms of $F(R)$ gravity. 

\subsection{$F(R)$ gravity}

The action of $F(R)$ gravity is written as 
\be
\label{JGRG7}
S_{F(R)}= \int d^4 x \sqrt{-g} \left( \frac{F(R)}{2\kappa^2} + \mathcal{L}_\mathrm{matter} \right)\, ,
\ee
with $g$ the determinant of the metric tensor $g_{\mu\nu}$. 
The equation of motion for modified gravity is given by
\be
\label{JGRG13}
\frac{1}{2}g_{\mu\nu} F(R) - R_{\mu\nu} F'(R) - g_{\mu\nu} \Box F'(R)
+ \nabla_\mu \nabla_\nu F'(R) = - \frac{\kappa^2}{2}T_{\mathrm{matter}\, \mu\nu}\, .
\ee
Here, 
${\nabla}_{\mu}$ is the covariant derivative operator and  
$\Box \equiv g^{\mu \nu} {\nabla}_{\mu} {\nabla}_{\nu}$
is the covariant d'Alembertian. 
Moreover, $T_{\mathrm{matter}\, \mu\nu}$ 
is the energy-momentum tensor of matter. 
By assuming spatially flat FLRW universe, 
the FLRW-like equations have the following forms:
\begin{align}
\label{JGRG15}
0 =& -\frac{F(R)}{2} + 3\left(H^2  + \dot H\right) F'(R)
 - 18 \left( 4H^2 \dot H + H \ddot H\right) F''(R) 
+ \kappa^2 \rho_\mathrm{matter}\, ,\\
\label{Cr4b}
0 =& \frac{F(R)}{2} - \left(\dot H + 3H^2\right)F'(R) 
+ 6 \left( 8H^2 \dot H + 4 {\dot H}^2 + 6 H \ddot H + \dddot H \right) F''(R) 
+ 36\left( 4H\dot H + \ddot H\right)^2 F'''(R) 
+ \kappa^2 P_\mathrm{matter}\, , 
\end{align}
where $\rho_\mathrm{matter}$ and $P_\mathrm{matter}$ are 
the energy density and pressure of matter, respectively, 
and the scalar curvature $R$ is given by $R=12H^2 + 6\dot H$. 
We note that $F'(R) \equiv dF(R)/dR$, while $R'(N) \equiv dR/dN$. 
{}From Eq.~(\ref{JGRG15}), we have 
\be
\label{SS9}
H^2 = \frac{ -F(R) + R F'(R)}{6 \left( F'(R) + R'(N) F''(R) \right)}\, .
\ee
In the following, to make mathematical expressions simpler, 
we omit the arguments of $H(N)$, $R(N)$, and $F(R)$. 
We again remark that the prime operating $H$ and $R$ means the 
derivative with respect to $N$, while the one operating $F$ 
denotes that with respect to $R$. 
Therefore, all of the slow-roll parameters
can be expressed in terms of $R$, $R'$, $R''$, $R'''$, and $F(R)$. 
Eventually, we acquire the representations of observables for inflationary 
models, $n_\mathrm{s}$, $r$, and $\alpha_\mathrm{s}$, 
presented in Appendix B.

\subsection{Reconstruction of $F(R)$ gravity models} 

The procedure of reconstruction of $F(R)$ gravity has been proposed in Ref.~\cite{NO-NOS}. 
We define the number of $e$-folds as $N \equiv \ln \left(a/a_0 \right)$ 
with $a_0$ the scale factor at the present time $t_0$, 
and the redshift becomes $z \equiv a_0/a - 1$. 
Accordingly, we have $N = -\ln \left( 1+z \right)$. 
We write the Hubble parameter in terms of $N$ via the function $G(N)$ as
\be
\label{RZ7}
\left(H(N)\right)^2=G(N) = G\left(- \ln\left(1+z\right)\right)\, .
\ee
Moreover, the scalar curvature is given by $R= 3 G'(N) + 12 G(N)$. 
For Eq.~(\ref{RZ7}), Eq.~(\ref{JGRG15}) is rewritten to 
\bea
0 \Eqn{=} 
-9 G\left(N\left(R\right)\right)\left(4 G'\left(N\left(R\right)\right)
+ G''\left(N\left(R\right)\right)\right) \frac{d^2 F(R)}{dR^2}
+ \left( 3 G\left(N\left(R\right)\right)
+ \frac{3}{2} G'\left(N\left(R\right)\right) \right) \frac{dF(R)}{dR} \nn
&& - \frac{F(R)}{2}
+ \sum_i \rho_{\mathrm{matter} \, i 0} a_0^{-3(1+w_i)} \exp \left[-3(1+w_i)N(R)\right]\, , 
\label{FRNe2}
\eea
where the last term in the right-hand side is equal to 
the total matter density $\rho_\mathrm{matter}$. 
Here, the matters are supposed to be fluids labeled 
by the subscription ``$i$'' 
with their constant equations of state 
$w_i \equiv P_{\mathrm{matter} \, i}/ \rho_{\mathrm{matter} \, i}$ 
with $\rho_{\mathrm{matter} \, i}$ and $P_{\mathrm{matter} \, i}$ 
the energy density and pressure of the $i$-th fluid, respectively, 
and $\rho_{\mathrm{matter} \, i 0}$ is a constant. 
In deriving Eq.~(\ref{FRNe2}), 
we have used the continuity equation $\dot{\rho}_i+3H(1+w_i)\rho_i=0$. 
Hence, by assuming a specific form of the Hubble parameter, i.e., $G(N)$, and using the relation between the Ricci scalar and the number of $e$-folds, Eq.~(\ref{FRNe2}) may be solved, so that the gravitational action $F(R)$ reproducing 
the expansion history described by $H$ can be found. 
On the other hand, for a particular $F(R)$ gravity model, the corresponding Hubble parameter $G(N)$ can be obtained. Thus, with the expressions for the observables of inflationary models derived in Sec.~III A, we may acquire the corresponding predictions on inflation for a particular action of $F(R)$ gravity.

\subsubsection{Linear form} 

Let us explore the linear form for $H^2$ in Eq.~(\ref{eq:IIC1}). 
The number of $e$-folds can be represented in terms of the Ricci scalar as 
$N(R)=\left(-3 G_0-12 G_1+R\right)/\left(12 G_0\right)$. 
The Friedmann equation in vacuum becomes a second order differential equation for $F(R)$ with respect to the Ricci scalar and 
its solution is derived as 
\be
F(R)=C_1(6G_0-2 R)^{3/2} \sqrt{\frac{R}{12
  G_0}-\frac{1}{4}} \left[1-\frac{1}{\frac{R}{12G_0}-\frac{1}{4}}-\frac{1}{4 \left(\frac{R}{12
  G_0}-\frac{1}{4}\right)^2}\right] +C_2 (6G_0-2 R)^{3/2} \text{L}\left(\frac{1}{2},\frac{3}{2};\frac{R}{12G_0}-\frac{1}{4}\right)\, . 
\label{FRNe6}
\ee
Here, $C_1$ and $C_2$ are integration constants, while $L(u_1,u_2;y)$ 
is the generalized Laguerre polynomial, 
where $u_1$ and $u_2$ are constants and $y$ is a variable. 
As a consequence, the corresponding gravitational Lagrangian 
has been reconstructed. 
The representations of the observables for inflationary models are written in Appendix C. 

The inflationary phase has to last long enough to account for the initial conditions problems, e.g., the so-called horizon and flatness problems. 
The value of number of $e$-folds at the end of inflation should be 
$N_\mathrm{e} \gtrsim 50$. 
The slow-roll parameters 
should be much smaller than unity 
during inflation, whereas 
the slow-roll parameters become larger than or 
equal to unity at the end of inflation, $N=N_\mathrm{e}$. 
As examples, if $(N, G_0, G_1) = (50.0, -0.850, 95.0)$ and 
$(60.0, -0.950, 115)$, we obtain 
$(n_\mathrm{s}, r, \alpha_\mathrm{s}) = (0.967, 0.121, -5.42 \times 10^{-5})$ 
and $(0.967, 0.123, -5.55 \times 10^{-5})$, respectively. 
Thus, the Planck results for $n_\mathrm{s}$ with $r = \mathcal{O}(0.1)$ 
can be realized. For illustrative purposes, we may fit the free parameters $\{G_0,G_1\}$ with the values of $(n_\mathrm{s}, r, \alpha_\mathrm{s})$ suggested by the Planck satellite and BICEP2 experiment by using the technique of the maximum likelihood, which is determined by the probability distribution:
\be
P(G_i) = {\cal N} \exp \left( -\frac{1}{2} \chi^2(G_i) \right)\,, 
\quad \text{where} 
\quad 
\chi^2 (G_i) \equiv \sum_{j}\frac{(y^{\mathrm{obs}}_{j} - y^{\mathrm{th}}_{j}(G_i))^2} {\sigma_{y^{obs}}^2}\,,
\label{Chi2}
\ee
with ${\cal N}$ a normalization factor and $G_i$ the free parameters of 
the model. 
Here, $\{y^\mathrm{obs}_j,  y^\mathrm{th}_{j}(G_i)\}$ are the observational data and the theoretical values predicted by a particular model respectively, while $\sigma_{y^\mathrm{obs}}$ are the errors. Supposing the number of $e$-folds $N_\mathrm{e}=50$, the best fit is given by the minimum of the function $\chi^2(G_i)$. For the model (\ref{FRNe6}), this gives $\chi^2_\mathrm{min}=2.322$, which corresponds to the set of free parameters $(G_0, G_1) = (3.03, -297.98)$, and lead to the following inflationary parameters:
\be
n_\mathrm{s}=0.958\,, \quad r=0.16\,, \quad  \alpha_\mathrm{s}=-0.00088\,.
\label{Result1} 
\ee 
  
\begin{figure}[t]
  \centering
\includegraphics[width=0.6\textwidth]{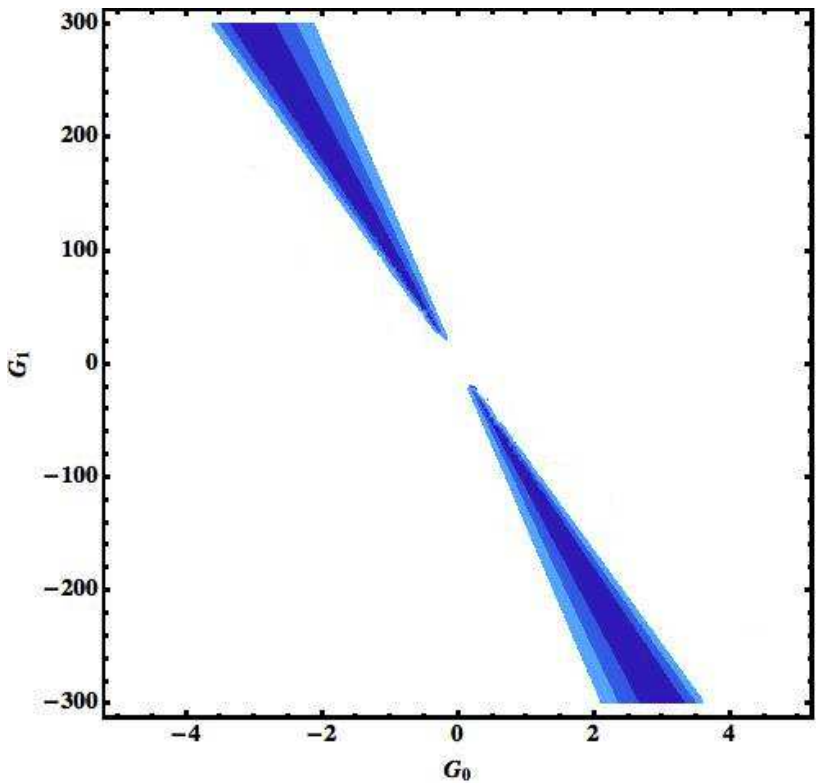}
\caption{Contour plot of the parameters $G_0$ and $G_1$ for the $F(R)$ gravity model in Eq.~(\ref{FRNe6}). From darker blue to lighter one, the contour shows 
1 $\sigma$, 2 $\sigma$, 2.58 $\sigma$, and 4 $\sigma$ 
confidence levels. 
} \label{fig1}
\end{figure}

The corresponding contour plot is depicted in Fig.~\ref{fig1}, which presents large errors and correlations in both parameters as expected, because the fit is provided by using just one more observational data than the number of free parameters. The contour displays 
1 $\sigma$, 2 $\sigma$, 2.58 $\sigma$, and 4 $\sigma$ 
confidence levels from darker blue to lighter one. 

We mention that the maximum likelihood leads to a set of free parameters $\{G_0,G_1\}$ that give approximately the correct results for 
the observables of inflationary models $(n_\mathrm{s}, r, \alpha_\mathrm{s})$. As a result, the inflationary era can be realized by the $F(R)$ gravity model with the form in Eq.~(\ref{FRNe6}). 

\subsubsection{Exponential form} 

Next, we investigate the exponential form for $H^2$ in Eq.~(\ref{eq:IIC5}). 
In this case, the relation between $N$ and $R$ is written as 
$\e^{\beta N}=\left(R-12G_3\right)/\left[3G_2 \left(4+\beta\right)\right]$. 
The solution for the Friedmann equation in vacuum becomes 
\bea
F(R)=C_1 F\left(b_{+},b_{-},l;y\right)+C_2\left(12G_3-R\right)^{\left(1+1/\beta \right)} F\left(1+b_{-}+\frac{1}{\alpha},1+b_{+}+\frac{1}{\alpha},2-d;y\right)
\, , 
\label{FRNe12}
\eea
with 
\bea
b_{\pm} = \frac{-3 \beta -2\pm\sqrt{\beta^2-20 \beta +4}}{4 \beta}\, , \quad 
d = -\frac{1}{\beta}\ , \quad y = \frac{12G_3-R}{12G_3+3G_3\beta}\, , 
\label{FRNe12-2}
\eea
where $F(v_1,v_2,v_3;y)$ with $v_j$ ($j=1, \dots, 3$) constants 
is the hypergeometric function. 

\begin{figure}[t]
  \centering
\includegraphics[width=0.6\textwidth]{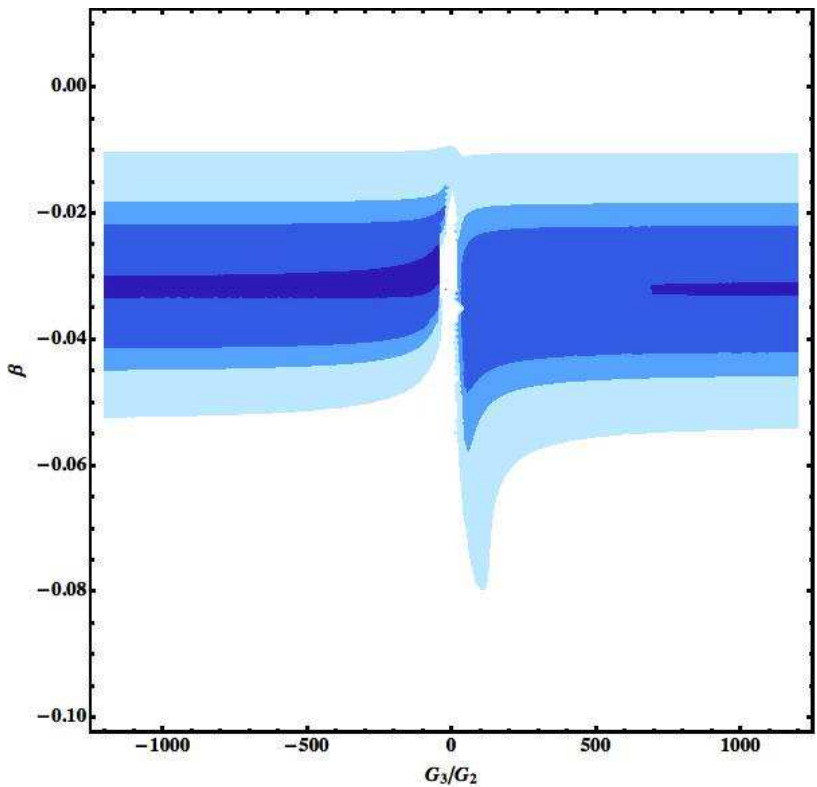}
\caption{Contour plot of the parameters $\beta$ and $G_3/G_2$ for the $F(R)$ gravity model in Eq.~(\ref{FRNe12}). 
Legend is the same as Fig.~\ref{fig1}.} \label{fig2}
\end{figure}

It follows from these expressions that 
for $(N, G_2, G_3) = (50.0, -1.10, 10.0)$ and 
$(60.0, -1.20, 15.0)$, we acquire 
$(n_\mathrm{s}, r, \alpha_\mathrm{s}) = (0.963, 6.89 \times 10^{-2}, -5.06 \times 10^{-5})$ 
and $(0.965, 5.84 \times 10^{-2}, -4.51 \times 10^{-5})$, respectively. 
Hence, these results are compatible with the Planck data on 
$n_\mathrm{s}$ and $r$. In addition, by fitting the free parameters with the Planck and BICEP2 data as done in the previous model, the maximum likelihood is given by $\chi^2_\mathrm{min}=2.72$, which corresponds to the free parameters $(\beta,G_2/G_3)=(-0.013, -4.04)$. Note that we have rewritten the Hubble parameter as $G(N)=G_2 \left[\e^{\beta N}+\left(G_3/G_2\right)\right]$, so that we can constrain $G_3/G_2$ in order to keep a less number of free parameters than number of data. 
Consequently, the best fit suggests the following values of 
the observables of inflationary models:
\be
n_\mathrm{s}=0.961\,, \quad r=0.20\,, \quad  \alpha_\mathrm{s}=0.0013\,. 
\label{Result2} 
\ee 
As in the previous model, the contour plot at the Fig.~\ref{fig2} shows large errors, particularly for $G_3/G_2$, whereas $\beta$ is much well constrained. However, the best fit presents quite good values for $n_\mathrm{s}$ and $r$ in spite of the deviation in $\alpha_\mathrm{s}$. 

%

\subsection{Power-law model of $F(R)$ gravity} 

Now, we take a concrete $F(R)$ gravity model 
and derive the observables of inflationary models.  
We study the power-law $F(R)$ gravity model of 
the action in Eq.~(\ref{JGRG7}) with 
$F(R) = \gamma R^n$, 
where $\gamma$ and $n$ are constants. 
In this model, the Friedmann equation (\ref{FRNe2}) for pure $F(R)$ gravity 
is a non-linear differential equation for $G(N)$. 
This has several independent solutions, one of which is given by
\be
G(N) = H^2 = G_4\e^{-2N/\zeta}\, , \quad \text{where}  \quad  
\zeta = \frac{-1+3n-2n^2}{n-2}\, , 
\label{FRNe17}
\ee
with $G_4 (>0)$ a positive constant. 
This form is equivalent to that in Eq.~(\ref{eq:IIC5}) with 
$G_2 = G_4$, $G_3 = 0$, and $\beta= -2/\zeta$. 
This solution is valid for $n\neq 2$. 
By using Eq.~(\ref{FRNe17}), 
we find that the slow-roll parameters are represented as 
$\epsilon = 1/\zeta$, 
$\eta = 2/\zeta$, and 
$\xi^2 = -4/\zeta$. 
Since the slow-roll parameters are constants, 
if $\zeta \gg 1$, we have $\epsilon \ll 1$, $\eta \ll 1$, and $\xi^2 \ll 1$ 
during inflation. 
There is an additional solution that may lead to 
the slow-roll regime and the following end of inflation. 
The Hubble parameter for the solution reads 
\be
G(N)=G_4 \exp\left[\frac{(-2+n)N+n(n-1)\ln \left(1-\exp\left\{-\frac{(-5+4n)
\left[N-2G_4n(n-1)\right]}{n-1}-3n+2n^2\right\}\right)}{(n-1)(2n-1)}\right]\, .
\label{FRNe19}
\ee

\begin{figure}[t]
\centering
\includegraphics[width=0.6\textwidth]{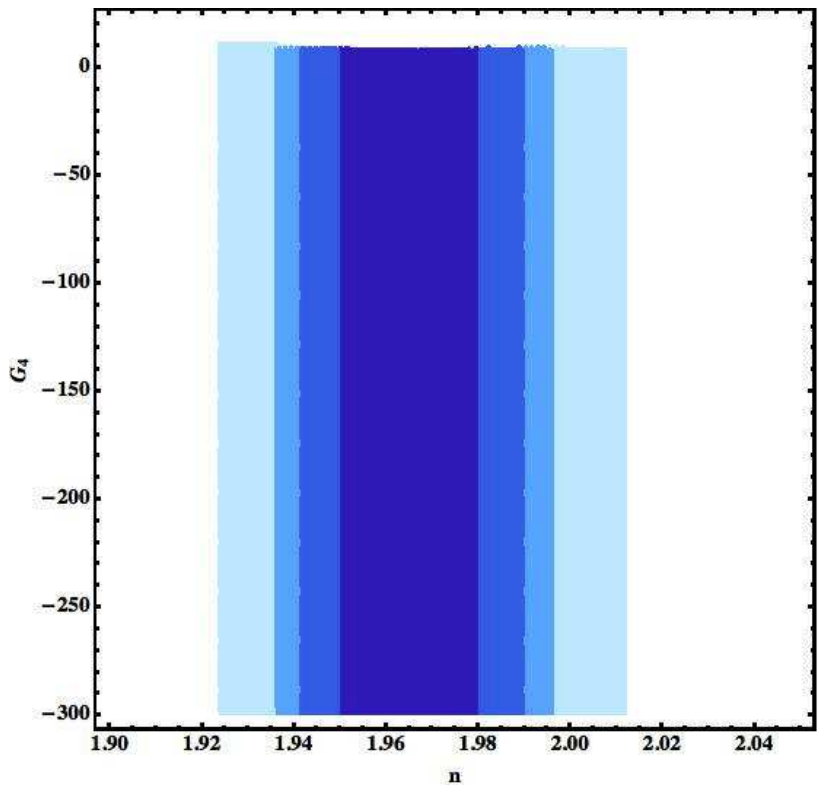}
\caption{Contour plot of the parameters $n$ and $G_4$ for the $F(R)$ gravity model of $F(R) = \gamma R^n$. 
Legend is the same as Fig.~\ref{fig1}.} \label{fig3}
\end{figure}

By a particular choice of the free parameters, one may be able to reproduce the correct observational values of $n_\mathrm{s}$ and/or $r$. By fitting the Hubble parameter (\ref{FRNe19}) with the data, we get $\chi^2_\mathrm{min}=2.25$ that corresponds to $(n,G_4)=(1.96,<8.84)$. Note that $G_4$ plays an important role, while $G_4$ in Eq.~(\ref{FRNe19}) is irrelevant for the values of the inflationary parameters. The inflationary parameters at the best fit read
\be
 n_\mathrm{s}=0.976\ , \quad r=0.18\ , \quad  \alpha_\mathrm{s}=-4.33\times 10^{-19}\ .
  \label{Result3} 
  \ee 
As shown in Fig.~\ref{fig3}, the value of $n$ is well constrained, while the initial condition $G_4$ allows a large range of the values.

We remark that 
the power-law model of $F(R) = \gamma R^n$ can be regarded as a particular limit of a viable $F(R)$ gravity model, 
which recovers General Relativity on small scales. 
For instance, in Ref.~\cite{Nojiri:2007cq} 
there has proposed the following model:  
$F(R)=R+\left[R^m \left(q_1 R^m - q_2 \right)
\right]/\left(1 + q_3 R^m\right)$ 
with $m$ and $q_j$ ($j=1, \dots, 3$) constants. 
At the inflationary stage, where the curvature is expected to be very large, 
this model tends to the power-law one, so that the above analysis can be applied for this particular viable $F(R)$ gravity model.

\section{Conclusions}

In the present paper, we have reconstructed the descriptions of inflation 
in perfect fluid models and $F(R)$ gravity theories. 
Particularly, as the observables of inflationary models, 
we have explored the spectral index of curvature perturbations, 
the tensor-to-scalar ratio, and the running of the spectral index. 
In addition, we have compared the perfect fluid and $F(R)$ gravity 
descriptions with the Planck and BICEP2 results. 

In the perfect fluid description, if the equation of state for the 
perfect fluid is approximately equal to minus unity, 
namely, the difference of the value between the equation of state and the 
cosmological constant is much smaller than unity, 
the Planck result for the spectral index of curvature perturbations with 
the tensor-to-scalar ratio of order of $0.1$ can be reproduced. 

On the other hand, in the description of $F(R)$ gravity, if the squared of the Hubble parameter has a linear form in terms of the number of $e$-folds during inflation, 
the spectral index is consistent with the Planck data and the 
tensor-to-scalar ratio becomes $\mathcal{O}(0.1)$, where as 
the Hubble parameter is given by an exponential function in terms of 
the number of $e$-folds, the value of the spectral index and 
the tensor-to-scalar ratio is compatible with the Planck analysis. 
By fitting the above $F(R)$ gravity models with the observational data released by Planck satellite and BICEP2 experiment, we have found that the best fit corresponds to the third model, $F(R)\propto R^n$, which provides the maximum likelihood according to the value of $\chi_\mathrm{min}^2$. Regarding this model, the predicted values for $n_\mathrm{s}$ and $r$ are within the allowed ranges suggested by the Planck and BICEP2 data. 
In spite of the great uncertainty of the integration constant $G_4$ and therefore on the initial conditions for the Hubble parameter, it provides a good constraint on the value of $n=1.96^{+0.02}_{-0.01}$.

It is finally remarked that 
the reconstruction method developed in this work 
may easily be extended for the case that there exists convenient matter. 
The corresponding analysis should not lead to any significant qualitative 
change in comparison with the analysis for pure gravity. 
It will be executed elsewhere.

\section*{Acknowledgments}

We would like to thank Professor Jaume de Haro for 
valuable discussions and suggestions very much. 
This work was partially supported by MINECO (Spain) project FIS2010-15640 (S.D.O. and D.S.-G.), the JSPS Grant-in-Aid 
for Scientific Research (S) \# 22224003 and (C) \# 23540296 (S.N.), 
that for Young Scientists (B) \# 25800136 (K.B.), 
the University of the Basque Country, Project Consolider CPAN Bo. CSD2007-00042 and the NRF financial support from the University of Cape Town (South Africa) 
(D.S.-G.).

\appendix

\section{Expressions of observables of inflationary models for the description of the perfect fluid}

In this Appendix, we present 
the observables of inflationary models in 
the description of the perfect fluid. 
Those expressions become 
\begin{eqnarray*} 
n_\mathrm{s} \Eqn{\sim} 1 - 9 \rho(N) f(\rho) 
\left( \frac{f'(\rho)-2}
{2\rho(N) - f(\rho)}\right)^2 
+\frac{6\rho(N)}{2\rho(N) -f(\rho)} 
\left\{ \frac{f(\rho)}{\rho(N)} + \frac{1}{2} 
\left(f'(\rho)\right)^2 + f'(\rho) 
-\frac{5}{2} \frac{f(\rho)f'(\rho)}{\rho(N)} + 
\left(\frac{f(\rho)}{\rho(N)}\right)^2
\right. \nonumber \\
&& \left. 
{}+\frac{1}{3} \frac{\rho'(N)}{f(\rho)} 
\left[\left(f'(\rho)\right)^2 + f(\rho) f''(\rho) 
-2 \frac{f(\rho)f'(\rho)}{\rho(N)} + \left( \frac{f(\rho)}{\rho(N)} 
\right)^2 \right] 
\right\} \,,
\label{eq:2.32} \\
r \Eqn{=} 24\rho(N) f(\rho) 
\left( \frac{f'(\rho)-2}
{2\rho(N) - f(\rho)}\right)^2 \,, 
\label{eq:2.33} \\
\alpha_\mathrm{s} \Eqn{=} \rho(N) f(\rho) \left( \frac{f'(\rho)-2}
{2\rho(N) - f(\rho)}\right)^2 \left[ 
\frac{72\rho(N)}{2\rho(N) - f(\rho)} J_1 
-54 \rho(N) f(\rho) \left( \frac{f'(\rho)-2}
{2\rho(N) - f(\rho)}\right)^2 
-\frac{1}{f'(\rho)-2}J_2
\right] \,,
\label{eq:2.34}
\end{eqnarray*}
where 
\begin{eqnarray*} 
J_1 \Eqn{\equiv} 
\frac{f(\rho)}{\rho(N)} + \frac{1}{2} 
\left(f'(\rho)\right)^2 + f'(\rho) 
-\frac{5}{2} \frac{f(\rho)f'(\rho)}{\rho(N)} + 
\left(\frac{f(\rho)}{\rho(N)}\right)^2
\nonumber \\
&& 
{}
+\frac{1}{3} \frac{\rho'(N)}{f(\rho)} 
\left[\left(f'(\rho)\right)^2 + f(\rho) f''(\rho) 
-2 \frac{f(\rho)f'(\rho)}{\rho(N)} + \left( \frac{f(\rho)}{\rho(N)} 
\right)^2 \right]\,, 
\label{eq:2.35} \\
J_2 \Eqn{\equiv} 
\frac{45}{2} \frac{f(\rho)}{\rho(N)} \left(f'(\rho)-\frac{1}{2} \frac{f(\rho)}{\rho(N)}\right) + 18\left(\frac{f(\rho)}{\rho(N)}\right)^{-1} 
\left(f'(\rho)-\frac{1}{2} \frac{f(\rho)}{\rho(N)}\right)^2 
+18\left(\frac{f(\rho)}{\rho(N)}\right)^{-1} 
\left(f'(\rho)-\frac{1}{2} \frac{f(\rho)}{\rho(N)}\right)^3 
\nonumber \\
&& 
{}
-9\left(f'(\rho)-\frac{1}{2} \frac{f(\rho)}{\rho(N)}\right)^2 
-45f'(\rho) + 9\frac{f(\rho)}{\rho(N)} 
\nonumber \\
&&
{}+3 \left(4f'(\rho) -7\frac{f(\rho)}{\rho(N)} +2\right) 
\left\{
-\frac{3}{2}\left(f'(\rho) -\frac{1}{2}\frac{f(\rho)}{\rho(N)}\right) 
\right. 
\nonumber \\
&& \left. 
{}+ \left(\frac{f(\rho)}{\rho(N)}\right)^{-2} 
\frac{\rho'(N)}{\rho(N)} 
\left[ \left(f'(\rho)\right)^2 + f(\rho) f''(\rho) 
-2 \frac{f(\rho)f'(\rho)}{\rho(N)} + \left( \frac{f(\rho)}{\rho(N)} 
\right)^2 \right] \right\} 
\nonumber \\
&&
{}+\left(\frac{f(\rho)}{\rho(N)}\right)^{-2} 
\left\{ 
-\frac{3}{2} \left(\frac{f(\rho)}{\rho(N)}\right) 
\left(\frac{\rho'(N)}{\rho(N)}\right) 
\left[ 
3\left( f'(\rho) \right)^2 
+2f(\rho) f''(\rho) 
-\frac{11}{2}\frac{f(\rho) f'(\rho)}{\rho(N)} 
+\frac{5}{2} \left( \frac{f(\rho)}{\rho(N)} \right)^2
\right]
\right. 
\nonumber \\ 
&& \left. 
{}
+ \left(\frac{\rho''(N)}{\rho(N)}\right) 
\left[
\left( f'(\rho) \right)^2 + f(\rho) f''(\rho) 
-2\frac{f(\rho) f'(\rho)}{\rho(N)} 
+ \left( \frac{f(\rho)}{\rho(N)} \right)^2
\right] 
\right. 
\nonumber \\ 
&& \left. 
{}
+ \left(\frac{\rho'(N)}{\rho(N)}\right)^2 
\left[ \left(3f'(\rho)f''(\rho) + f(\rho)f'''(\rho)  \right) \rho(N) 
-3\left( f'(\rho) \right)^2 - 3f(\rho) f''(\rho) 
\right. 
\right. 
\nonumber \\ 
&& \left. \left. 
{}
+6\frac{f(\rho) f'(\rho)}{\rho(N)} 
-3\left( \frac{f(\rho)}{\rho(N)} \right)^2 
\right] 
\right\} 
\,. 
\label{eq:2.36}
\end{eqnarray*}

\section{Expressions of observables of inflationary models for the $F(R)$ gravity description}

In this Appendix, we write 
the observables of inflationary models in the $F(R)$ gravity description. 
The representations read
\begin{eqnarray*} 
n_\mathrm{s} \Eqn{\sim} 1 
-\frac{3 \left(F-R F'\right) \left(F'' R' \left(R'+2 R\right)+F' \left(R'-2 R\right)-4
   F\right)^2}{2 \left(-R F'' R'-2 R F'+F\right)^2 \left(R F'' R'-R F'+2
   F\right)}
\nonumber \\  
&&
{}-\frac{1}{2
   \left(F-R F'\right)^2 \left(R F'' R'-R F'+2 F\right)^2 \left(R F'' R'+2 R
   F'-F\right)}
\left(
\left(R F'-F\right) 
\right. 
\nonumber \\  
&&
\left.
{}\times 
\left(\left(R F'' R'-R F'+2 F\right)^4-18 
\left(F-RF'\right) \left(R F'' R'-R F'+2 F\right)^3
\right. 
\right.
\nonumber \\  
&&
\left.
\left.
{}
+\left(11 F^2+R \left(6 R \left(F''\right)^2
   \left(R'\right)^2-F' \left(R F'+4 F\right)+6 F'' R' \left(3 F-R F'\right)\right)\right)
\right.
\right. 
\nonumber \\  
&&
\left.
\left.
\left(-F'' \left(R'\right)^2 \left(R^2 F''+F\right)+4 F^2+R \left(F'\right)^2
   \left(R'+3 R\right)-F' \left(R' \left(R F'' \left(2 R-R'\right)+F\right)+8 F
   R\right)\right)
\right. 
\right.
\nonumber \\  
&&
\left.
\left.
{}
+2 \left(R F'-F\right) \left(R' \left(F'' R'+F'\right) \left(-2 R F''
   R'-3 R F'+2 F\right)
\right.
\right. 
\right.
\nonumber \\  
&&
\left.
\left.
\left.
{}
-\frac{\left(R F'' R'-R F'+2 F\right) \left(4 F^2+R \left(3 R
   \left(F''\right)^2 \left(R'\right)^2+7 R \left(F'\right)^2+F' \left(6 R F'' R'-8
   F\right)\right)\right)}{F-R F'}
\right.
\right. 
\right.
\nonumber \\  
&&
\left.
\left.
\left.
{}
+\left(R F'-F\right) \left(F''' \left(R'\right)^3+F''
   R' \left(2 R''+R'\right)+F' R''\right)\right) \left(R F'' R'-R F'+2 F\right)\right)
\right) \,,
\\
r \Eqn{=} - \frac{4\left(-F + R F'\right)}{2 F - R F' + R R' F''} 
\left[ \frac{ R' F' + \left(R'\right)^2 F'' - 4F - 2 R F' + 2 R R' F'' }{ -F + 2R F' + R R' F''}\right]^2 \,, 
\\
\alpha_\mathrm{s} \Eqn{=} 
\frac{1}{2 \left(F-R F'\right)^2
   \left(F-2 R F'-R R' F''\right)^4
   \left(2 F-R F'+R R'
   F''\right)^4}
\left(
\left(R F'-F\right) 
\right.
\nonumber \\  
&&
\left.
{}\times 
\left(3
   \left(F-R F'\right)^3 \left(2
   F-R F'+R R' F''\right)^2
   \left(-4 F+F' \left(R'-2
   R\right)+R' \left(2 R+R'\right)
   F''\right)^4
\right. 
\right.
\nonumber \\  
&&
\left.
\left.
{}
+2 \left(R
   F'-F\right) \left(2 F-R F'+R R'
   F''\right) \left(-F+2 R F'+R R'
   F''\right) \left(\left(2 F-R
   F'+R R' F''\right)^4
\right.
\right. 
\right.
\nonumber \\  
&&
\left.
\left.
\left.
{}
-18
   \left(F-R F'\right) \left(2 F-R
   F'+R R' F''\right)^3+2 \left(R
   F'-F\right) \left(R' \left(F'+R'
   F''\right) \left(2 F-3 R F'-2 R
   R' F''\right)
\right.
\right.
\right. 
\right.
\nonumber \\  
&&
\left.
\left.
\left.
\left.
{}
-\frac{\left(2 F-R
   F'+R R' F''\right) \left(4 F^2+R
   \left(7 R
   \left(F'\right)^2+\left(6 R R'
   F''-8 F\right) F'+3 R
   \left(R'\right)^2
   \left(F''\right)^2\right)\right)
   }{F-R F'}
\right.
\right. 
\right.
\right.
\nonumber \\  
&&
\left.
\left.
\left.
\left.
{}
+\left(R F'-F\right)
   \left(F'''
   \left(R'\right)^3+F'' \left(R'+2
   R''\right) R'+F'
   R''\right)\right) \left(2 F-R
   F'+R R' F''\right)
\right. 
\right.
\right.
\nonumber \\  
&&
\left.
\left.
\left.
{}
+\left(11
   F^2+R \left(6 R
   \left(R'\right)^2
   \left(F''\right)^2+6 \left(3 F-R
   F'\right) R' F''-F' \left(4 F+R
   F'\right)\right)\right) \left(4
   F^2+R \left(F'\right)^2 \left(3
   R+R'\right)
\right. 
\right. 
\right.
\right.
\nonumber \\  
&&
\left.
\left.
\left.
\left.
{}
-\left(R'\right)^2
   F'' \left(F'' R^2+F\right)-F'
   \left(8 F R+R' \left(F+R \left(2
   R-R'\right)
   F''\right)\right)\right)\right)
   \left(-4 F+F' \left(R'-2
   R\right)+R' \left(2 R+R'\right)
   F''\right)^2 
\right. 
\right. 
\nonumber \\  
&&
\left.
\left.
{}
+\left(F-R F'\right)
   \left(F-2 R F'-R R' F''\right)^2
   \left(4 F+F' \left(R'-2
   R\right)+R' \left(2 R+R'\right)
   F''\right) \left(256 F^5
\right. 
\right.
\right.
\nonumber \\  
&&
\left.
\left.
\left.
{}
+R^2
   \left(F'\right)^5
   \left(\left(R'\right)^3+6 R
   \left(R'\right)^2+2 R \left(3
   R+R''\right) R'+R^2 \left(-16
   R+6 R''+R'''\right)\right)
\right. 
\right. 
\right. 
\nonumber \\  
&&
\left.
\left.
\left.
{}
+R
   \left(F'\right)^4 \left(R
   \left(F'''' R^2+2 F''' R+3
   F''\right)
   \left(R'\right)^4+\left(9
   F''' R^3+8 F'' R^2-2 F\right)
   \left(R'\right)^3
\right. 
\right. 
\right. 
\right. 
\nonumber \\  
&&
\left.
\left.
\left.
\left.
{}
+R \left(R
   \left(5 R R'' F'''-2 F''
   \left(R-2 R''\right)\right)-23
   F\right) \left(R'\right)^2+2 R
   \left(R^2 F'' \left(32 R+7
   R''\right)-F \left(17 R+4
   R''\right)\right) R'
\right. 
\right. 
\right. 
\right. 
\nonumber \\  
&&
\left.
\left.
\left.
\left.
{}
+3 F R^2
   \left(48 R-9 R''-2
   R'''\right)\right)+R'
   \left(-8 R^4 \left(R'\right)^5
   \left(F''\right)^5
\right. 
\right. 
\right. 
\right. 
\nonumber \\  
&&
\left.
\left.
\left.
\left.
{}
+F R^2
   \left(R'\right)^3 \left(-2
   \left(R'\right)^2-39 R R'+4 R
   \left(4 R-5 R''\right)\right)
   \left(F''\right)^4
\right. 
\right. 
\right. 
\right. 
\nonumber \\  
&&
\left.
\left.
\left.
\left.
{}
+F
   \left(R'\right)^2
   \left(\left(F-11 R^3
   F'''\right)
   \left(R'\right)^3-13 F R
   \left(R'\right)^2-F R \left(53
   R+4 R''\right) R'+F R^2
   \left(128 R-83 R''+2
   R'''\right)\right)
   \left(F''\right)^3
\right. 
\right. 
\right. 
\right. 
\nonumber \\  
&&
\left.
\left.
\left.
\left.
{}
+F^2 R'
   \left(R \left(R F''''-2
   F'''\right)
   \left(R'\right)^4-46 R^2 F'''
   \left(R'\right)^3+\left(5 R^2
   R'' F'''-18 F\right)
   \left(R'\right)^2-8 F R'' R'
\right. 
\right. 
\right. 
\right. 
\right. 
\nonumber \\  
&&
\left.
\left.
\left.
\left.
\left.
{}
+4 F
   R \left(96 R-23 R''+2
   R'''\right)\right)
   \left(F''\right)^2+4 F^3
   \left(\left(R
   F''''-F'''\right)
   \left(R'\right)^4-13 R F'''
   \left(R'\right)^3+5 R R''
   F''' \left(R'\right)^2
\right. 
\right. 
\right. 
\right. 
\right. 
\nonumber \\  
&&
\left.
\left.
\left.
\left.
\left.
{}
+7 F
   R'+F \left(128 R-3 R''+2
   R'''\right)\right) F''+4 F^4
   R' \left(F''''
   \left(R'\right)^2-2 F''' R'+5
   R'' F'''\right)\right)
\right. 
\right. 
\right. 
\nonumber \\  
&&
\left.
\left.
\left.
{}
-F'
   \left(R \left(F''\right)^2
   \left(-2 R^2
   \left(F''\right)^2+\left(2 F-11
   R^3 F'''\right) F''+2 F R
   \left(R F''''-2
   F'''\right)\right)
   \left(R'\right)^6
\right. 
\right. 
\right. 
\right. 
\nonumber \\  
&&
\left.
\left.
\left.
\left.
{}
-F'' \left(13
   \left(F''\right)^3 R^4+21 F
   \left(F''\right)^2 R^2+2 F^2
   \left(4 F'''-5 R
   F''''\right) R+F F'' \left(59
   F''' R^3+3 F\right)\right)
   \left(R'\right)^5
\right. 
\right. 
\right. 
\right. 
\nonumber \\  
&&
\left.
\left.
\left.
\left.
{}
+\left(4 R^4
   \left(4 R-5 R''\right)
   \left(F''\right)^4-F R^2
   \left(19 R+8 R''\right)
   \left(F''\right)^3-5 F R \left(5
   F-2 R^2 R'' F'''\right)
   \left(F''\right)^2-64 F^2 R^2
   F''' F''
\right. 
\right. 
\right. 
\right. 
\right. 
\nonumber \\  
&&
\left.
\left.
\left.
\left.
\left.
{}
+4 F^3
   \left(F'''+3 R
   F''''\right)\right)
   \left(R'\right)^4+F \left(R^3
   \left(192 R-97 R''+4
   R'''\right)
   \left(F''\right)^3+2 F R
   \left(25 R-7 R''\right)
   \left(F''\right)^2
\right. 
\right. 
\right. 
\right. 
\right. 
\nonumber \\  
&&
\left.
\left.
\left.
\left.
\left.
{}
+2 F \left(25
   R'' F''' R^2+14 F\right)
   F''+20 F^2 R F'''\right)
   \left(R'\right)^3+F^2 \left(R^2
   \left(768 R-71 R''+19
   R'''\right)
   \left(F''\right)^2
\right. 
\right. 
\right. 
\right. 
\right. 
\nonumber \\  
&&
\left.
\left.
\left.
\left.
\left.
{}
+4 F \left(14
   R+3 R''\right) F''+60 F R R''
   F'''\right)
   \left(R'\right)^2+4 F^3 \left(R
   F'' \left(320 R+23 R''+5
   R'''\right)-8 F\right) R'
\right. 
\right. 
\right. 
\right. 
\nonumber \\  
&&
\left.
\left.
\left.
\left.
{}
+4
   F^4 \left(192 R+3
   R''-R'''\right)\right)-\left(
   F'\right)^3 \left(R^2 F''
   \left(2 R^2 F''''-3 F''\right)
   \left(R'\right)^5
\right. 
\right. 
\right. 
\right. 
\nonumber \\  
&&
\left.
\left.
\left.
\left.
{}
+R \left(8 R^2
   \left(F''\right)^2+\left(7
   F''' R^3+6 F\right) F''+2 F R
   \left(4 F'''+3 R
   F''''\right)\right)
   \left(R'\right)^4
\right. 
\right. 
\right. 
\right. 
\nonumber \\  
&&
\left.
\left.
\left.
\left.
{}
+\left(R^2
   \left(F'' \left(45 F+R F''
   \left(17 R+2 R''\right)\right)+R
   \left(43 F+10 R F'' R''\right)
   F'''\right)-F^2\right)
   \left(R'\right)^3
\right. 
\right. 
\right. 
\right. 
\nonumber \\  
&&
\left.
\left.
\left.
\left.
{}
+R
   \left(\left(F''\right)^2
   \left(96 R+26 R''+3
   R'''\right) R^3+F F''
   \left(11 R+20 R''\right) R+3 F
   \left(10 R^2 R'' F'''-9
   F\right)\right)
   \left(R'\right)^2
\right. 
\right. 
\right. 
\right. 
\nonumber \\  
&&
\left.
\left.
\left.
\left.
{}
+F R \left(R^2
   F'' \left(448 R+97 R''+4
   R'''\right)-2 F \left(36 R+5
   R''\right)\right) R'+F^2 R^2
   \left(512 R-33 R''-13
   R'''\right)\right)
\right. 
\right. 
\right. 
\nonumber \\  
&&
\left.
\left.
\left.
{}
+\left(F'
\right)^2 \left(R^2
   \left(F''\right)^2 \left(F''+R
   \left(R F''''-2
   F'''\right)\right)
   \left(R'\right)^6-R F'' \left(8
   R^2 \left(F''\right)^2+\left(13
   F''' R^3+6 F\right) F''
\right. 
\right. 
\right. 
\right. 
\right. 
\nonumber \\  
&&
\left.
\left.
\left.
\left.
\left.
{}
+4 F R
   \left(F'''-2 R
   F''''\right)\right)
   \left(R'\right)^5+\left(4 R^3
   \left(2 R-R''\right)
   \left(F''\right)^3+R^2 \left(5
   R'' F''' R^2+F\right)
   \left(F''\right)^2
\right. 
\right. 
\right. 
\right. 
\right. 
\nonumber \\  
&&
\left.
\left.
\left.
\left.
\left.
{}
+F \left(3 F-5
   R^3 F'''\right) F''+F^2 R
   \left(10 F'''+13 R
   F''''\right)\right)
   \left(R'\right)^4+R \left(2 R^3
   \left(32 R-7 R''+R'''\right)
   \left(F''\right)^3
\right. 
\right. 
\right. 
\right. 
\right. 
\nonumber \\  
&&
\left.
\left.
\left.
\left.
\left.
{}
+F R \left(65
   R-4 R''\right)
   \left(F''\right)^2+5 F \left(8
   R'' F''' R^2+13 F\right)
   F''+62 F^2 R F'''\right)
   \left(R'\right)^3
\right. 
\right. 
\right. 
\right. 
\nonumber \\  
&&
\left.
\left.
\left.
\left.
{}
+F
   \left(\left(F''\right)^2
   \left(480 R+47 R''+14
   R'''\right) R^3+F F''
   \left(51 R+28 R''\right) R+5 F
   \left(13 R^2 R'' F'''-2
   F\right)\right)
   \left(R'\right)^2
\right. 
\right. 
\right. 
\right. 
\nonumber \\  
&&
\left.
\left.
\left.
\left.
{}
+F^2 \left(R^2
   F'' \left(187 R''+16 \left(72
   R+R'''\right)\right)-4 F
   \left(18 R+R''\right)\right)
   R'+4 F^3 R \left(224 R-3
   R'''\right)\right)\right)\right)
\right)
\,.
\end{eqnarray*}
%

\section{Linear and exponential forms of the Hubble parameter}

In this Appendix, we write the expressions of the observables for inflationary models in the linear form of $H^2$ in Eq.~(\ref{eq:IIC1}) 
and those in the exponential form of $H^2$ in Eq.~(\ref{eq:IIC5}). 
For the linear form, we find 
\begin{eqnarray*}
n_\mathrm{s} \Eqn{=} \frac{36\left(G_0 N + G_1 \right)^2 + 84 G_0 \left(G_0 N + G_1 \right) - 5G_0^2}{\left[ 6\left(G_0 N + G_1 \right) + G_0 \right]^2}\, , 
\nonumber \\
r \Eqn{=} -\frac{268G_0 \left(G_0 N + G_1 \right)}{\left[ 6\left(G_0 N + G_1 \right) + G_0 \right]^2}\, , 
\nonumber \\ 
\alpha_\mathrm{s} \Eqn{=} - \frac{27 G_0^2}{2 \left(G_0 N + G_1 \right)^2 \left[ 6\left(G_0 N + G_1 \right) + G_0 \right]^4} \left\{ 
64 \left(G_0 N + G_1 \right)^3 
\left[ 3\left(G_0 N + G_1 \right) - G_0 \right]
\right.
\nonumber \\  
&&
\left. 
{}-3G_0 \left[ 2\left(G_0 N + G_1 \right) - G_0 \right] \left[ 6\left(G_0 N + G_1 \right) + G_0 \right]^2 
\right\}\, . 
\end{eqnarray*}

On the other hand, for the exponential form, we obtain 
\begin{eqnarray*}
n_\mathrm{s} \Eqn{=} 
\frac{-6\left(\beta^2 +6\beta -6\right) G_3^2 
+2 \left( \beta^3 +6\beta^2 +6\beta +36\right) G_3G_2 \e^{\beta N}
+\left( \beta^3 +13\beta^2 +48\beta +36\right) G_2^2 \e^{2\beta N}}{\left[6G_3+(6+\beta)G_2\e^{\beta N}\right]^2}\,,  
\nonumber \\
r \Eqn{=} 
-\frac{8\beta \left(6+\beta \right)^2 G_2 \e^{\beta N} 
\left(G_3+G_2\e^{\beta N}\right)}{\left[6G_3+(6+\beta)G_2 
\e^{\beta N} \right]^2}\, ,  
\nonumber \\
\alpha_\mathrm{s} \Eqn{=} 
\frac{\beta^2 \left(6+\beta \right)^2 G_2 \e^{\beta N}}{\left[6G_3+(6+\beta)G_2\e^{\beta N} \right]^4} 
\left\{ 4\left(6+\beta \right) \left(G_3+G_2\e^{\beta N}\right) \left(G_3+2G_2\e^{\beta N}\right)\left[6G_3+(6+\beta)G_2\e^{\beta N} \right] 
\right. \nonumber \\ 
&&
\left. 
{}-6\left(6+\beta \right)^2 G_2 \e^{\beta N} 
\left(G_3+G_2\e^{\beta N}\right)^2 
-\frac{1}{2} G_2\e^{\beta N} \left[6G_3+(6+\beta)G_2\e^{\beta N} \right]^2 
\left[ -\frac{3}{4}\beta^2 \left( \frac{G_2 \e^{\beta N}}{G_3+G_2\e^{\beta N}} \right)^2 
\right. 
\right. 
\nonumber \\ 
&&
\left. 
\left. 
{}
+\frac{3}{2} \beta \left(1+\beta-\beta^2\right) 
\left( \frac{G_2 \e^{\beta N}}{G_3+G_2\e^{\beta N}} \right) 
+\left(-9+\frac{15}{2}\beta+\beta^2+6\beta^3 \right) 
\right. 
\right. 
\nonumber \\ 
&&
\left. 
\left. 
{}
+ 6\beta\left(1-\beta^2 \right) \left( \frac{G_2 \e^{\beta N}}{G_3+G_2\e^{\beta N}} \right)^{-1} \right] 
\right\} \,.  
\end{eqnarray*}
%

\section{Relation between the EoS parameter 
and the tensor-to-scalar ratio}

In this Appendix, we explore the relation between the EoS parameter 
and the tensor-to-scalar ratio. 
In Ref.~\cite{Vazquez:2013dva},  
Eq.~(1.3) gives an expression for the tensor-to-scalar ratio:
\be
r(k) = 64\pi \left( \frac{{\dot \phi}^2}{H^2} \right)_{k=aH}\,, 
\tag{D1}
\ee
where a unit of $G=1$ has maybe been used. 
Then, Eq.~(D1) can be rewritten as
\be
r(k) = 8\kappa^2 \left( \frac{{\dot \phi}^2}{H^2} \right)_{k=aH}\, . 
\tag{D2}
\ee
For the scalar field, the EoS parameter $w$ is given by
\be
w = \frac{P}{\rho}\, ,\quad 
P = \frac{1}{2}{\dot \phi}^2 - V(\phi)\, , \quad 
\rho = \frac{1}{2}{\dot \phi}^2 + V(\phi)\, .
\tag{D3}
\ee

By using the Friedmann equation in the FLRW background 
\be
H^2 = \frac{\kappa^2}{3} \rho\, ,
\tag{D4}
\ee
we find
\be
V(\phi) = \left( \frac{24}{r(k)} - \frac{1}{2} \right) {\dot\phi}^2\, ,
\tag{D5}
\ee
and therefore
\be
H^2 = \frac{\kappa^2}{3} \frac{24}{r(k)} {\dot\phi}^2\, .
\tag{D6}
\ee
Thus, $w$ can be expressed in term of $r(k)$ as follows, 
\be
w = - 1 + \frac{r(k)}{24}\, .
\tag{D7}
\ee

\end{document}